# A Fixed-Parameter Algorithm for #SAT with Parameter Incidence Treewidth[*]

Marko Samer and Stefan Szeider

Department of Computer Science
University of Durham, UK
{marko.samer,stefan.szeider}@durham.ac.uk

**Abstract.** We present an efficient fixed-parameter algorithm for #SAT parameterized by the incidence treewidth, i.e., the treewidth of the bipartite graph whose vertices are the variables and clauses of the given CNF formula; a variable and a clause are joined by an edge if and only if the variable occurs in the clause. Our algorithm runs in time $\mathcal{O}(4^k\, k\, l\, N)$, where $k$ denotes the incidence treewidth, $l$ denotes the size of a largest clause, and $N$ denotes the number of nodes of the tree-decomposition.

## 1 Introduction

The counting problem #SAT is the problem of computing the number of satisfying truth assignments of a given propositional formula in conjunctive normal form (CNF). Several problems in automatic reasoning and artificial intelligence can be naturally encoded as #SAT problem. However, since the problem is #P-complete [18], it is very unlikely that it can be solved in polynomial time. Moreover, #SAT remains #P-hard even for monotone 2CNF formulas and Horn 2CNF formulas, and it is NP-hard to approximate the number of satisfying assignments of a CNF formula with $n$ variables within an error of $2^{n^{1-\varepsilon}}$ for $\varepsilon > 0$. This approximation hardness holds also for monotone 2CNF formulas and Horn 2CNF formulas [15].

The structure of a CNF formula can be naturally encoded by various graph concepts. The most popular one is called the *primal graph* whose vertices are variables and two variables are joined by an edge if and only if they occur together in some clause. Symmetrically, the vertices of the *dual graph* are clauses and two clauses are joined by an edge if and only if there is a variable which occurs in both of them. The most important graph concept in this paper, however, is the *incidence graph*, a bipartite graph with two disjoint sets of vertices: the first one consists of the clauses and the second one consists of the variables; there is an edge joining a clause in the first set and a variable in the second set if and only if the variable occurs in the clause. See Figure 1 for an illustration of these concepts. It is easy to see that all these graphs encode some structural information of the CNF formula. The important observation now is that this structural information allows us in many cases to count the number of satisfying assignments more efficiently. Note that it is unlikely that the satisfiability of a CNF formula with $n$ variables can be decided in time $2^{o(n)}$ [10]. Hence, it is expected that in the worst case we cannot count the number of satisfying assignments significantly faster than by considering all $2^n$ possible assignments. However, for a given instance, the structural information tells us which parts of the given CNF formula can be processed independently. So in certain cases we can compute the number of satisfying assignments significantly more efficiently than by exhaustive search. To this aim, we consider a *tree-decomposition* of the graph; each tree-decomposition has an associated *width*. The smaller the width of the tree-decomposition, the less parts of the given CNF formula depend on each other and consequently the more efficiently the corresponding instance can be solved. The *treewidth* of a graph is the minimum width over all such tree-decompositions.

---

[*] Research supported by the EPSRC project EP/E001394/1.

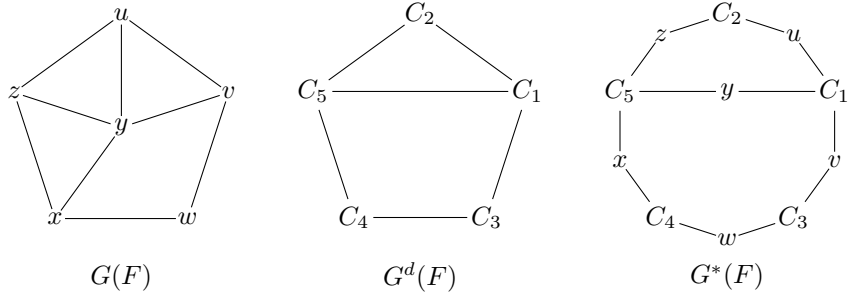

**Fig. 1.** Graphs associated with the CNF formula $F = \{C_1, \ldots, C_5\}$ with $C_1 = \{u, \neg v, \neg y\}$, $C_2 = \{\neg u, z\}$, $C_3 = \{v, \neg w\}$, $C_4 = \{w, \neg x\}$, $C_5 = \{x, y, \neg z\}$; the primal graph $G(F)$, the dual graph $G^d(F)$, and the incidence graph $G^*(F)$.

In this paper we present an efficient fixed-parameter algorithm for #SAT parameterized by the incidence treewidth $tw^*$. The concept of fixed-parameter algorithms was introduced by Downey and Fellows within their theory of parameterized complexity [5]. In particular, a fixed-parameter algorithm solves instances of size $n$ and parameter $k$ in time $\mathcal{O}(f(k)\,n^c)$, where $f$ denotes a computable function and $c$ denotes a constant that is independent of parameter $k$. A fixed-parameter algorithm remains feasible for large instances provided that the parameter $k$ is small even if the function $f$ is exponential. This feature makes fixed-parameter algorithms often preferable to algorithms with runtime $\mathcal{O}(n^{f(k)})$, since the latter become infeasible for large instances even if $f$ is linear and $k$ is small. For an in-depth treatment of parameterized complexity we refer the reader to other sources [5, 7, 12].

Note that the *existence* of a fixed-parameter algorithm for #SAT parameterized by the incidence treewidth follows already from monadic second order theory [4, 9, 17]; however, an algorithm obtained in this way is practically infeasible. Our algorithm, on the other hand, is practically feasible. Assume that a CNF formula $F$ and a tree-decomposition of width $k$ are given; let $l$ denote the size of a largest clause of $F$ and $N$ denote the number of nodes of the tree-decomposition. We present an algorithm that computes the number of satisfying assignments in time $\mathcal{O}(4^k\,k\,l\,N)$; there are no big hidden constants. Note that $N$ is linear in the sum of the number of variables and the number of clauses.

As an immediate consequence of our algorithm, we obtain also fixed-parameter algorithms with respect to the primal treewidth $tw$, since $tw^* \leq tw + 1$ [11], the dual treewidth $tw^d$, since $tw^* \leq tw^d + 1$ [11], and the branchwidth $bw$, since $tw \leq 3/2\, bw$ [14] and thus $tw^* \leq 3/2\, bw + 1$. Branchwidth is a measure analogous to treewidth for a decomposition method called *branch-decomposition*. Note that in addition to the above inequalities, $tw$, $tw^d$, and $bw$ are strictly less general than $tw^*$ [9, 17]. A fixed-parameter algorithm for SAT with respect to primal treewidth has previously been published by Gottlob et al. [8], and a fixed-parameter algorithm for #SAT with respect to branchwidth has previously been published by Bacchus et al. [1]. Fischer et al. [6] propose a parameterized algorithm for #SAT with respect to the parameter incidence treewidth; it appears to us that important details of their algorithm are let out, making it difficult to verify the claimed time complexity.

The various treewidth parameters can be defined analogously for instances of the constraint satisfaction problem (CSP), considering constraints instead of clauses. From the work of Gottlob et al. [8] it follows that the Boolean CSP is fixed-parameter tractable with respect to the parameter primal treewidth. In contrast to SAT and #SAT, this result cannot be generalized to the more general parameter incidence treewidth (subject to a complexity theoretic assumption): Samer and Szeider [16] have shown that the Boolean CSP parameterized by the incidence treewidth is W[1]-hard.

A different approach for solving #SAT was presented by Nishimura, Ragde, and Szeider [13]. They developed a fixed-parameter algorithm for computing strong backdoor sets with respect to cluster formulas, which yields a fixed-parameter algorithm for #SAT. In terms of generality, the corresponding parameter *clustering-width* is incomparable with incidence treewidth.

## 2 Preliminaries

### 2.1 Tree-Decompositions

Let $G$ be a graph, $T$ a tree, and $\chi$ a labeling of the vertices of $T$ by sets of vertices of $G$. We refer to the vertices of $T$ as "nodes" to avoid confusion with the vertices of $G$. The pair $(T, \chi)$ is a *tree-decomposition* of $G$ if the following three conditions hold:

1. For every vertex $v \in V(G)$ there exists a node $t \in V(T)$ such that $v \in \chi(t)$.
2. For every edge $\{v, w\} \in E(G)$ there exists a node $t \in V(T)$ such that $v, w \in \chi(t)$.
3. For any three nodes $t_1, t_2, t_3 \in V(T)$, if $t_2$ lies on a path from $t_1$ to $t_3$, then $\chi(t_1) \cap \chi(t_3) \subseteq \chi(t_2)$ ("Connectedness Condition").

The *width* of a tree-decomposition $(T, \chi)$ is defined by $\max_{t \in V(T)} |\chi(t)| - 1$. The *treewidth* $tw(G)$ of a graph $G$ is the minimum width over all its tree-decompositions.

For constant $k$, there exists a linear time algorithm that checks whether a given graph has treewidth at most $k$ and, if so, outputs a tree-decomposition of minimum width [2].

For our purposes it is convenient to consider a special type of tree-decompositions. Let $(T, \chi)$ be a tree-decomposition of a graph $G$ and let $r$ be a node of $T$. The triple $(T, \chi, r)$ is a *nice tree-decomposition* of $G$ if the following three conditions hold; here we consider $T$ as a tree rooted at $r$:

1. Every node of $T$ has at most two children.
2. If a node $t$ of $T$ has two children $t_1$ and $t_2$, then $\chi(t) = \chi(t_1) = \chi(t_2)$; in that case we call $t$ a *join node*.
3. If a node $t$ of $T$ has exactly one child $t'$, then exactly one of the following prevails:
   (a) $|\chi(t)| = |\chi(t')| + 1$ and $\chi(t') \subset \chi(t)$; in that case we call $t$ an *introduce node*.
   (b) $|\chi(t)| = |\chi(t')| - 1$ and $\chi(t) \subset \chi(t')$; in that case we call $t$ a *forget node*.

It is well known that one can transform any tree-decomposition of width $k$ in linear time into a nice tree-decomposition of width at most $k$ [3].

Let $(T, \chi, r)$ be a nice tree-decomposition of a graph $G$. For each node $t$ of $T$ let $T_t$ denote the subtree of $T$ rooted at $t$; furthermore, let $G_t$ denote the subgraph of $G$ which is induced by the vertex set $V_t = \bigcup_{t' \in V(T_t)} \chi(t')$. Observe that $(T_t, \chi|_{V(T_t)}, t)$ is a nice tree-decomposition of $G_t$.

### 2.2 Propositional Satisfiability and #SAT

We consider propositional formulas $F$ in conjunctive normal form (CNF) represented as sets of clauses. Each clause in $F$ is a finite set of *literals*, and a literal is a negated or unnegated propositional *variable*. For example,

$$F = \{\{\neg x, y, z\}, \{\neg y, \neg z\}, \{x, \neg y\}\}$$

represents the propositional formula $(\neg x \vee y \vee z) \wedge (\neg y \vee \neg z) \wedge (x \vee \neg y)$. For a clause $C$ we denote by $var(C)$ the set of variables that occur (negated or unnegated) in $C$; for a formula $F$ we put $var(F) = \bigcup_{C \in F} var(C)$. The *size* of a clause is its cardinality.

A *truth assignment* is a mapping $\tau : X \to \{0, 1\}$ defined on some set $X$ of variables. We extend $\tau$ to literals by setting $\tau(\neg x) = 1 - \tau(x)$ for $x \in X$. A truth assignment $\tau : X \to \{0, 1\}$ *satisfies* a clause $C$ if for some variable $x \in var(C) \cap X$ we have $x \in C$ and $\tau(x) = 1$ or $\neg x \in C$ and $\tau(x) = 0$. An assignment satisfies a formula $F$ if it satisfies all clauses in $F$. A formula $F$ is *satisfiable* if there exists a truth assignment that satisfies $F$; otherwise $F$ is *unsatisfiable*. For a formula $F$ we denote by $\#(F)$ the number of assignments $\tau : var(F) \to \{0, 1\}$ that satisfy $F$. Thus $F$ is satisfiable if and only if $\#(F) \geq 1$. The *propositional satisfiability problem* SAT is the problem of deciding whether a given propositional formula in CNF is satisfiable. The *counting SAT problem* #SAT is the problem of computing $\#(F)$ of a given propositional formula $F$ in CNF.

### 2.3 Parameterization

A *satisfiability parameter* is a computable function $p$ that assigns to every CNF formula $F$ a non-negative integer $p(F)$. The parameterized problem instances consist of a CNF formula $F$ and a non-negative integer $k$ with $p(F) \leq k$. We will use the satisfiability parameter incidence treewidth $tw^*$ as described in the following.

The *incidence graph* $G^*(F)$ of a CNF formula $F$ is the bipartite graph with vertex set $F \cup var(F)$; a variable $x$ and a clause $C$ are joined by an edge if and only if $x \in var(C)$. The *incidence treewidth* $tw^*(F)$ of a CNF formula $F$ is the treewidth of its incidence graph, that is $tw^*(F) = tw(G^*(F))$.

## 3 The Fixed-Parameter Algorithm

For this section, let $(T, \chi, r)$ be a nice tree-decomposition of the incidence graph $G^*(F)$ of a CNF formula $F$. Let $k$ denote the width of $(T, \chi, r)$.

For each node $t$ of $T$, let $F_t$ denote the set consisting of all the clauses in $V_t$, and let $X_t$ denote the set of all variables in $V_t$, i.e., $F_t = V_t \cap F$ and $X_t = V_t \cap var(F)$. We also use the shorthands $\chi_c(t) = \chi(t) \cap F$ and $\chi_v(t) = \chi(t) \cap var(F)$ for the set of variables and the set of clauses in $\chi(t)$, respectively.

Let $t$ be a node of $T$. For each truth assignment $\alpha : \chi_v(t) \to \{0,1\}$ and a subset $A \subseteq \chi_c(t)$ we define $N(t, \alpha, A)$ as the set of truth assignments $\tau : X_t \to \{0,1\}$ for which the following two conditions hold:

1. $\tau(x) = \alpha(x)$ for all variables $x \in \chi_v(t)$.
2. $A$ is exactly the set of clauses of $F_t$ that are not satisfied by $\tau$.

We represent the values of $n(t, \alpha, A) = |N(t, \alpha, A)|$ for all $\alpha$ and $A$ by a table $M_t$ with $|\chi(t)| + 1$ columns and $2^{|\chi(t)|}$ rows. The first $|\chi(t)|$ columns of $M_t$ contain Boolean values encoding $\alpha(x)$ for variables $x \in \chi_v(t)$, and membership of $C$ in $A$ for clauses $C \in \chi_c(t)$. We denote the row of $M_t$ that encodes $\alpha$ and $A$ by $M_t(\alpha, A)$. The last entry of each row $M_t(\alpha, A)$ contains the integer $n(t, \alpha, A)$.

**Lemma 1** *Let $t$ be a* join node *of $T$ with children $t_1, t_2$. Then, for each assignment $\alpha : \chi_v(t) \to \{0,1\}$ and set $A \subseteq \chi_c(t)$, we have*

$$n(t, \alpha, A) = \sum_{A_1, A_2 \subseteq \chi_c(t),\ A_1 \cap A_2 = A} n(t_1, \alpha, A_1) \cdot n(t_2, \alpha, A_2).$$

*Proof.* In the following, we will show that the mapping $f : \tau \mapsto (\tau|_{X_{t_1}}, \tau|_{X_{t_2}})$ is a bijection from the set $N(t, \alpha, A)$ into the set $M = \{(\tau_1, \tau_2) \mid \text{there exists } A_1 \subseteq \chi_c(t_1) \text{ and } A_2 \subseteq \chi_c(t_2) \text{ with } A_1 \cap A_2 = A \text{ such that } \tau_1 \in N(t_1, \alpha, A_1) \text{ and } \tau_2 \in N(t_2, \alpha, A_2)\}$. The above equality follows then immediately.

First, let us show that $f$ is a mapping from $N(t, \alpha, A)$ into $M$. To this aim, let $\tau \in N(t, \alpha, A)$ and $f(\tau) = (\tau_1, \tau_2)$. Now, let $A_1$ and $A_2$ be exactly the sets of clauses of $F_{t_1}$ and $F_{t_2}$ that are not satisfied by $\tau_1$ and $\tau_2$ respectively. Since $X_t = X_{t_1} \cup X_{t_2}$, we know that a clause is satisfied by $\tau$ if and only if it is satisfied by $\tau_1$ or $\tau_2$. Thus, since $F_t = F_{t_1} \cup F_{t_2}$, we have $A \subseteq A_1 \cap A_2$ and $A_1 \cap A_2 \subseteq A$, that is, $A_1 \cap A_2 = A$. In addition, we know that $A_1 \subseteq \chi_c(t_1)$. For the sake of contradiction, let us assume that there exists a clause $C \in F_{t_1} \setminus \chi_c(t_1) \subseteq F_t$ which is not satisfied by $\tau_1$. If $C$ is not satisfied by $\tau$, then $C \in A \subseteq \chi_c(t) = \chi_c(t_1)$, which contradicts our assumption. Otherwise, if $C$ is satisfied by $\tau$, then $C$ must also be satisfied by $\tau_2$, since it is not satisfied by $\tau_1$. Thus, there exists a variable $x \in X_{t_2}$ which occurs also in $C$ and satisfies $C$ under the assignment $\tau_2(x) = \tau(x)$. By the definition of a tree-decomposition of an incidence graph, however, this implies that $x \in X_{t_1}$. So we have $x \in X_{t_1} \cap X_{t_2}$. Note that, by the connectedness condition, we know that $V_{t_1} \cap V_{t_2} = \chi(t)$, that is, $X_{t_1} \cap X_{t_2} = \chi_v(t)$ and $F_{t_1} \cap F_{t_2} = \chi_c(t)$. Hence, it follows that $x \in \chi_v(t) = \chi_v(t_1)$, which implies that $\tau(x) = \tau_1(x)$. Thus, $C$ is satisfied by $\tau_1$, which again contradicts our assumption. So we have $A_1 \subseteq \chi_c(t_1)$. It is now easy to see that $\tau_1 \in N(t_1, \alpha, A_1)$. The case of $\tau_2$ is completely analogous. Consequently, $f$ is indeed a mapping from $N(t, \alpha, A)$ into $M$.

To show that $f$ is injective, let $\tau, \sigma \in N(t, \alpha, A)$ such that $f(\tau) = f(\sigma)$. Then, since $\tau|_{X_{t_1}} = \sigma|_{X_{t_1}}$ and $\tau|_{X_{t_2}} = \sigma|_{X_{t_2}}$, we know that $\tau = \sigma$. To show that $f$ is surjective, let $(\tau_1, \tau_2) \in M$. Now, let us define the assignment $\tau : X_t \to \{0, 1\}$ by $\tau|_{X_{t_1}} = \tau_1$ and $\tau|_{X_{t_2}} = \tau_2$. Since $X_t = X_{t_1} \cup X_{t_2}$, we know that a clause is satisfied by $\tau$ if and only if it is satisfied by $\tau_1$ or $\tau_2$. Thus, we know that $A = A_1 \cap A_2$ is exactly the set of clauses of $F_t = F_{t_1} \cup F_{t_2}$ that are not satisfied by $\tau$. It is now easy to see that $\tau \in N(t, \alpha, A)$. Consequently, $f$ is indeed a bijection from $N(t, \alpha, A)$ into $M$. □

**Lemma 2** *Let $t$ be an* introduce node *with child $t'$.*
(a) *If $\chi(t) = \chi(t') \cup \{x\}$ for a variable $x$, then, for each assignment $\alpha : \chi_v(t') \to \{0,1\}$ and set $A \subseteq \chi_c(t)$, we have*

$$n(t, \alpha \cup \{(x, 0)\}, A) = \begin{cases} 0 & \text{if } \neg x \in C \text{ for a clause } C \in A; \\ \sum_{B' \subseteq B} n(t', \alpha, A \cup B') & \text{otherwise, where } B = \{C \in \chi_c(t) \mid \neg x \in C\}; \end{cases}$$

$$n(t, \alpha \cup \{(x, 1)\}, A) = \begin{cases} 0 & \text{if } x \in C \text{ for some clause } C \in A; \\ \sum_{B' \subseteq B} n(t', \alpha, A \cup B') & \text{otherwise, where } B = \{C \in \chi_c(t) \mid x \in C\}. \end{cases}$$

(b) *If $\chi(t) = \chi(t') \cup \{C\}$ for a clause $C$, then, for each assignment $\alpha : \chi_v(t) \to \{0,1\}$ and clause set $A \subseteq \chi_c(t)$, we have*

$$n(t, \alpha, A) = \begin{cases} n(t', \alpha, A) & \text{if } C \notin A \text{ and } \alpha \text{ satisfies } C; \\ n(t', \alpha, A \setminus \{C\}) & \text{if } C \in A \text{ and } \alpha \text{ does not satisfy } C; \\ 0 & \text{otherwise.} \end{cases}$$

*Proof. (a)* Let us consider the case of $N(t, \alpha \cup \{(x, 0)\}, A)$; the case of $N(t, \alpha \cup \{(x, 1)\}, A)$ is completely symmetric. Note that by definition $N(t, \alpha \cup \{(x, 0)\}, A) = \emptyset$ if there is some clause $C$ in $A$ such that $C$ contains $\neg x$. Thus, let us assume that no clause in $A$ contains $\neg x$. Moreover, let $B = \{C \in \chi_c(t) \mid \neg x \in C\}$. In the following, we will show that the mapping $f : \tau \mapsto \tau|_{X_{t'}}$ is a bijection from the set $N(t, \alpha \cup \{(x, 0)\}, A)$ into the set $\bigcup_{B' \subseteq B} N(t', \alpha, A \cup B')$. Note that always $N(t', \alpha, A \cup B') \cap N(t', \alpha, A \cup B'') = \emptyset$ for $B' \neq B''$. The above equality follows then immediately.

For any $\tau \in N(t, \alpha \cup \{(x, 0)\}, A)$, let $f(\tau) = \tau'$. It is then easy to see that $\tau' \in N(t', \alpha, A \cup B')$ for some $B' \subseteq B$. To show that $f$ is injective, let $\tau, \sigma \in N(t, \alpha \cup \{(x, 0)\}, A)$ such that $f(\tau) = f(\sigma)$. Then, since $\tau|_{X_{t'}} = \sigma|_{X_{t'}}$ and $\tau(x) = \sigma(x) = 0$ for the single variable $x \in X_t \setminus X_{t'}$, we know that $\tau = \sigma$. To show that $f$ is surjective, let $\tau' \in N(t', \alpha, A \cup B')$ for some $B' \subseteq B$. Now we define the assignment $\tau : X_t \to \{0, 1\}$ by $\tau|_{X_{t'}} = \tau'$ and $\tau(x) = 0$. It is then easy to see that $\tau \in N(t, \alpha \cup \{(x, 0)\}, A)$. Consequently, $f$ is indeed a bijection from $N(t, \alpha \cup \{(x, 0)\}, A)$ into $\bigcup_{B' \subseteq B} N(t', \alpha, A \cup B')$.

*(b)* Note that by definition $N(t, \alpha, A) = \emptyset$ if $C \in A$ and $\alpha$ satisfies $C$ or $C \notin A$ and $\alpha$ does not satisfy $C$ for the single clause $C \in \chi_c(t) \setminus \chi_c(t')$. Thus, let us assume that (i) $C \notin A$ and $\alpha$ satisfies $C$ or (ii) $C \in A$ and $\alpha$ does not satisfy $C$. In the following, we will show that the mapping $f : \tau \mapsto \tau$ is a bijection from the set $N(t, \alpha, A)$ into the set $N(t', \alpha, A)$ in case (i) resp. into the set $N(t', \alpha, A \setminus \{C\})$ in case (ii). The above equalities follow then immediately.

For any $\tau \in N(t, \alpha, A)$, it is easy to see that $\tau \in N(t', \alpha, A)$ in case (i) and $\tau \in N(t', \alpha, A \setminus \{C\})$ in case (ii). Moreover, since $f(\tau) = \tau$, it follows trivially that $f$ is injective. To show that $f$ is surjective, let $\tau \in N(t', \alpha, A)$ in case (i) and $\tau \in N(t', \alpha, A \setminus \{C\})$ in case (ii). Under the assumption of case (i) resp. case (ii), it is then easy to see that $\tau \in N(t, \alpha, A)$. Consequently, $f$ is indeed a bijection from $N(t, \alpha, A)$ into $N(t', \alpha, A)$ in case (i) resp. into $N(t', \alpha, A \setminus \{C\})$ in case (ii). □

**Lemma 3** *Let $t$ be a* forget node *with child $t'$.*
(a) *If $\chi(t) = \chi(t') \setminus \{x\}$ for a variable $x$, then, for each assignment $\alpha : \chi_v(t) \to \{0, 1\}$ and set $A \subseteq \chi_c(t)$, we have*

$$n(t, \alpha, A) = n(t', \alpha \cup \{(x, 0)\}, A) + n(t', \alpha \cup \{(x, 1)\}, A).$$

(b) *If $\chi(t) = \chi(t') \setminus \{C\}$ for a clause $C$, then, for each assignment $\alpha : \chi_v(t) \to \{0, 1\}$ and set $A \subseteq \chi_c(t)$, we have*

$$n(t, \alpha, A) = n(t', \alpha, A).$$

*Proof.* (a) It is easy to see that the mapping $f : \tau \mapsto \tau$ is a bijection from the set $N(t, \alpha, A)$ into the set $N(t', \alpha \cup \{(x, 0)\}, A) \cup N(t', \alpha \cup \{(x, 1)\}, A)$. The above equality follows then immediately.

(b) It is easy to see that the mapping $f : \tau \mapsto \tau$ is a bijection from the set $N(t, \alpha, A)$ into the set $N(t', \alpha, A)$. The above equality follows then immediately. □

**Lemma 4** *Let $t$ be a* leaf node. *Then, for each assignment $\alpha : \chi_v(t) \to \{0, 1\}$ and set $A \subseteq \chi_c(t)$, we have*

$$n(t, \alpha, A) = \begin{cases} 1 & \textit{if } A = \{\, C \in \chi_c(t) \mid \alpha \textit{ does not satisfy } C \,\}; \\ 0 & \textit{otherwise.} \end{cases}$$

*Proof.* Since $X_t = \chi_v(t)$ and $F_t = \chi_c(t)$ for every leaf node $t$, we know that for each assignment $\tau : X_t \to \{0, 1\}$ there exists exactly one assignment $\alpha : \chi_v(t) \to \{0, 1\}$ (and vice versa) such that $\tau(x) = \alpha(x)$ for all variables $x \in X_t$. Hence, the above equivalence follows immediately. □

By using these equivalences, we can now construct the tables $M_t$ from the leaves to the root according to the following lemma. We assume that multiplication of integers can be carried out in time $\mathcal{O}(1)$. It is easy to adjust our results for other models of computation.

**Lemma 5** *Let $t$ be a node of $T$. Given the tables of the children of $t$, we can compute the table $M_t$ in time $\mathcal{O}(4^k \, k \, l)$, where $l$ is the size of a largest clause of $F$.*

*Proof.* To check the runtime of computing $M_t$, let $p = |\chi_v(t)|$ and $q = |\chi_c(t)|$; since we assume that the width of the tree-decomposition under consideration is $k$, we have $p + q \leq k + 1$. Now, let us distinguish between the different kinds of nodes.

(i) Let $t$ be a *join node* with children $t_1, t_2$. We compute the table $M_t$ from the tables $M_{t_1}$ and $M_{t_2}$ according to Lemma 1 as follows: First we initialize the last column of $M_t$ to 0. For each of the $2^p$ choices of $\alpha$, we consider all $2^q$ possibilities for $A_1$ and all $2^q$ possibilities for $A_2$; we increase the last entry of the row $M_t(\alpha, A_1 \cap A_2)$ by $n(t_1, \alpha, A_1) \cdot n(t_2, \alpha, A_2)$. Taking the intersection of $A_1$ and $A_2$ and updating the last entry of a row can be accomplished in time $\mathcal{O}(q)$. Hence, we can compute $M_t$ in time $\mathcal{O}(2^p\, 2^q\, 2^q\, q) \subseteq \mathcal{O}(4^k\, k\, l)$.

(ii) Let $t$ be an *introduce node* with child $t'$. We compute the table $M_t$ from table $M_{t'}$ according to Lemma 2 as follows: For each of the $2^p$ choices of $\alpha$, we consider all $2^q$ possibilities for $A$. In case (a), we set the last entry of the rows $M_t(\alpha' \cup \{(x,0)\}, A)$ and $M_t(\alpha' \cup \{(x,1)\}, A)$ to the last entry of the row $M_{t'}(\alpha', A)$ or to 0 depending on whether the literals $\neg x$ or $x$ occur in some clause in $A$. Checking whether a literal occurs in some clause in $A$ can be accomplished in time $\mathcal{O}(q\, l)$. Hence, we can compute $M_t$ in time $\mathcal{O}(2^p\, 2^q\, q\, l) \subseteq \mathcal{O}(4^k\, k\, l)$. In case (b), we set the last entry of the row $M_t(\alpha, A)$ to the last entry of the rows $M_{t'}(\alpha, A)$ or $M_{t'}(\alpha, A \setminus \{C\})$ or to 0 depending on whether $C \in A$ and $\alpha$ satisfies $C$. Checking whether $C \in A$ and $\alpha$ satisfies $C$ can be accomplished in time $\mathcal{O}(p\, l)$. Hence, we can compute $M_t$ in time $\mathcal{O}(2^p\, 2^q\, p\, l) \subseteq \mathcal{O}(4^k\, k\, l)$.

(iii) Let $t$ be a *forget node* with child $t'$. We compute the table $M_t$ from table $M_{t'}$ according to Lemma 3 as follows: For each of the $2^p$ choices of $\alpha$, we consider all $2^q$ possibilities for $A$. In case (a), we set the last entry of the row $M_t(\alpha, A)$ to the sum of the last entries of the rows $M_{t'}(\alpha \cup \{(x,0)\}, A)$ and $M_{t'}(\alpha \cup \{(x,1)\}, A)$. Hence, we can compute $M_t$ in time $\mathcal{O}(2^p\, 2^q) \subseteq \mathcal{O}(4^k\, k\, l)$. In case (b), we set the last entry of the row $M_t(\alpha, A)$ to the last entry of the row $M_{t'}(\alpha, A)$. Hence, we can compute $M_t$ in time $\mathcal{O}(2^p\, 2^q) \subseteq \mathcal{O}(4^k\, k\, l)$.

(iv) Let $t$ be a *leaf node*. We compute the table $M_t$ according to Lemma 4 as follows: For each of the $2^p$ choices of $\alpha$, we consider all $2^q$ possibilities for $A$; we set the last entry of the row $M_t(\alpha, A)$ to 1 or 0 depending on whether $A$ is equal to the subset of clauses of $\chi_c(t)$ that are not satisfied by $\alpha$. Checking whether $A$ is equal to the subset of clauses of $\chi_c(t)$ that are not satisfied by $\alpha$ can be accomplished in time $\mathcal{O}(p\, q\, l)$. Hence, we can compute table $M_t$ in time $\mathcal{O}(2^p\, 2^q\, p\, q\, l) \subseteq \mathcal{O}(4^k\, k\, l)$. $\square$

**Theorem 1** *Given a nice tree-decomposition of the incidence graph of a CNF formula $F$, then we can compute #($F$) in time $\mathcal{O}(4^k\, k\, l\, N)$; $k$ denotes the width, $l$ denotes the size of a largest clause, and $N$ denotes the number of nodes of the tree-decomposition. Consequently, #SAT parameterized by the incidence treewidth is fixed-parameter tractable.*

*Proof.* If a CNF formula $F$ and a non-negative integer $k$ are given, we can check in linear time whether $tw^*(F) \leq k$ and, if so, we can compute a nice tree-decomposition of minimal width (see Section 2.1). Let $(T, \chi, r)$ be a nice tree-decomposition of the incidence graph of $F$; let $k$ and $n$ be the width and number of nodes of $(T, \chi, r)$ respectively. Starting from the leaf nodes of $T$ we compute all $N$ tables $M_t$ for $t \in V(T)$ in a bottom up ordering. Each table can be computed by Lemma 5 in time $\mathcal{O}(4^k\, k\, l)$. Since we have

$$\#(F) = \sum_{\alpha: \chi_v(r) \to \{0,1\}} n(r, \alpha, \emptyset),$$

we can read off #($F$) from the table $M_r$ at the root node $r$. □